\documentclass[preprint,12pt]{elsarticle}
\usepackage[export]{adjustbox}
\usepackage{rotating}
\usepackage{pdflscape}
\usepackage{lineno,hyperref}

\usepackage{amssymb}
\usepackage{amsthm}
\usepackage[utf8x]{inputenc}
\usepackage{lineno}
\usepackage{graphicx}
\usepackage{subcaption}
\journal{Material Research Express}
\begin{document}
	
\begin{frontmatter}




\title{Lattice thermal conductivity of pristine and doped (B,N) Graphene }


\author{Sarita Mann$^a$, Isha Mudahar$^b$, Hitesh Sharma$^c$*, V.K. Jindal$^a$, Girija S. Dubey$^{d,e,f}$, Godfrey Gumbs $^f$ and Vassilios Fessatidis $^e$ }

\address[a]{Department of Physics, Panjab University, Chandigarh, India-160014}
\address[b]{Department of Basic and Applied Sciences, Punjabi University, Patiala, India-147002 }
\address[c]{Department of Physical Sciences, IK Gujral Punjab Technical University, Kapurthala, Punjab, India-146003}
\address[d] {Department of Earth and Physical Sciences, York College of CUNY, Jamaica, 11451}
\address[e] {Department of Physics and Engineering Physics, Fordham University, Bronx, New York, NY 10458}
\address[f] {Department of Physics and Astronomy, Hunter College, CUNY 695 Park Anvenue, New York, NY 10065}

\begin{abstract}
In this paper, the effect of B and N doping on the phonon induced thermal conductivity of graphene has been investigated. This study is important when one has to evaluate the usefulness of electronic properties of B and N doped graphene. We have performed the calculations by employing density functional perturbation theory(DFPT) to calculate the inter-atomic forces$/$force constants of pristine/doped graphene. Thermal conductivity calculations have been carried out by making use of linearized Boltzmann transport equations (LBTE) under single-mode relaxation time approximation(RTA). The thermal conductivity of pristine graphene has been found to be of the order of 4000W/mK at 100K, which decreases gradually with an increase in temperature. The thermal conductivity decreases drastically by 96  $\%$ to 190 W/mK when doped with 12.5 $\%$ B and reduces by 99 $\%$ to 30 W/mK with 25 $\%$ B doping. When graphene is doped with N, the thermal conductivity decreases to 4 W/mK and 55 W/mK for 12.5 $\%$ and 25 $\%$ doping concentration, respectively. We have found that the thermal conductivity of doped graphene show less sensitivity to change in temperature. It has also been shown that the thermal conductivity of graphene can be tuned with doping and has a strong dependence on doping concentration.
	
\end{abstract}

\end{frontmatter}


\section{Introduction}
From a technological point of view, heat removal from devices plays a key role in producing ultra-small and energy-efficient electronic, optoelectronic, and photovoltaic devices and systems \cite{r1,r101,r102}. In recent years, one of the key issues for making further progress has been performance enhancement in energy conversion and thermal management technologies, which can be implemented into engineering applications \cite{r103}. The choice of thermal interface materials (TIM) between the emitter and receiver is critical to determine the effectiveness of any potential device. In addition, the "green" revolution in a photovoltaic cell has also highlighted the need for novel thermally efficient materials based on graphene \cite{r103}. In low-dimensional structures, thermal conduction has disclosed many fascinating characteristics. When the dimension of a material is reduced to the nano dimension in electronic devices, there is a significant reduction in the thermal conductivity as a result of the enhanced boundary scattering, which is lower than the bulk value \cite{r2}. The wide range of thermal properties of carbon nanomaterials has highlighted the possibility of their usage in thermal management,heat dissipation and thermoelectric conversion\cite{r221,r91}.
\par
In the last decade, different theoretical methods and experimental techniques have been used for investigating thermal transport in carbon nanosystems \cite{r222}. 
At room temperature (RT), thermal conductivity of carbon nanomaterials have been found to vary from 0.01 W/mK in amorphous carbons, 600 W/mK in supported graphene\cite{r31}, 2300 W/mK in carbon nanotubes to more than 4000 W/mK in graphene \cite{r201,r202}.  Isolated studies have even reported higher thermal conductivity of 5000 W/mK at room temperature of perfect single-layer graphene prepared by mechanical exfoilation \cite{r3}. Further, with the increase in the temperature the thermal conductivity is found to decrease. Using optothermal Raman method, the thermal conductivity of suspended graphene samples decrease from (2500+1100/-1050) W/mK near 350 K to (1400+500/-480) W/mK at near 500 K \cite{r4}. The thermal conductivity has shown isotopic effect, with isotropically pure $^{12}C$  (0.01$\%$ $^{13}C$)  graphene, showing thermal conductivity of 4,000 W/ mK at 320K, and is found to decrease by a factor of two in graphene sheets made up of half-half ratio of $^{12}C$ and $^{13}C$ \cite{r401}.
\par

In addition to temperature dependence,it has also been found that the thermal conductance in graphene depends on layer thickness and application of the biaxial tensile strain. At room temperature, thermal conductivity of graphene has been found in the range of 1300-2800 W/mK when the number of graphene atomic planes were increased from 2 to 4[15].The thermal conductivity of graphene sheet has been found to decrease by applying biaxial tensile strain. Under a biaxial tensile strain of 0.12 $\%$, there is a drop of around 20 $\%$ in the thermal conductivity of suspended mono-layer graphene at about 350 K and approximately 12 $\%$ drop near 500 K. \cite{r51}.
\par
Therefore, it is of interest to understand the change of thermal conductivity of graphene due to other controllable factors such as the introduction of defects or doping and reducing the dimension of the graphene sheet. B and N have been investigated as a preferred dopant in graphene for doping due to their similar size to C atoms and low formation energy in graphene. N doped graphene has shown higher electrochemical activity than pristine graphene \cite{r11}, and B doped graphene has shown interesting optical or electrochemical properties \cite{r12}. Therefore, B/N doped graphene is an effective way to modify its electronic, chemical and thermal properties \cite{r10}. Therefore, understanding the thermal properties of doped graphene as a function of temperature is of fundamental scientific interest as a possible material for thermal management.
\par
Theoretically, thermal conductivity has been calculated using phonon properties obtained by solving lattice dynamics equations based on the Boltzmann transport equation (BTE) and non-equilibrium Green's function. Theoretical methods such as Kubo-Greenwood and non-equilibrium molecular dynamics have demonstrated their capability to calculate thermal conductivity directly with reasonable accuracy \cite{r6}. The use of empirical potentials has led to a deviation of the thermal conductivity from experimental results. In recent years, the calculation of the force constants using first-principle calculations using density functional theory (DFT) and density functional perturbation theory (DFPT), have demonstrated quantitative accuracy for describing a wide range of thermal conductivity \cite{r6,r61}. The ab initio calculations have reported high thermal conductivity (TC) of graphene in the wide range of the order 1900-4400 W/mK \cite{r62, r7, r71}, 3000 - 5000 W/mK \cite{r8} and 5400-8700 W/mK \cite{r9}. Therefore, it is of importance to calculate the accurately thermal conductivity of pristine graphene before investigating the effect of doping systematically on thermal conductivity.
\par
In our group, we have studied the structural and electronic properties of various carbon nanostructures \cite{r10,r1001,r1002,r1003}. In the present work, we have extended our investigation to study the phonon induced thermal conductivity of pristine and doped graphene by considering non-harmonic contributions using a combination of ab initio methods and Boltzmann transport equation. We have studied the effect of doping by B and N with a doping concentration of  12.5 $\%$ and 25 $\% $.
\section{Methodology and Computational details}

Thermal conductivity and phonon properties of graphene have been calculated using net force exerted on each atom within the foundation of density functional theory (DFT) as implemented in the Vienna ab-initio simulation package(VASP) software \cite{r11,r12}.
Interatomic force constants $\Phi_{\alpha \beta}$ were calculated using
\begin{equation}
\Phi_{\alpha \beta} (lk,l'k')= \frac{\partial^2\phi}{\partial r_{\alpha} (lk)\partial r_{\beta}(lk)} ,
\end{equation}
where \emph{l,k} are the positions of any atom w.r.t other atom at \emph{l'k'}. The Cartesian coordinates are represented by $\alpha$, $\beta$ and the position vector of the atom is denoted by \emph{r}.\par
To describe the interaction between ions and electrons,a projector augmented wave (PAW) approach is used in VASP calculation \cite{r13}. Generalized Gradient Approximation (GGA) is used to carry out the calculations which apply Perdew, Burke and Ernzerhof (PBE) exchange correlation \cite{r14}. 
The self consistent energy convergence was set to 10$^{-8}$ eV/atom. The Brilloiun zone integration was done on an 7x7x1 K mesh and a plane wave basis set of 750 eV cut off energy is used. Structure was relaxed untill all forces were equal to or less than 10$^{-4}$ eV/$\AA$. For performing the phonon dispersion calculation we have used the phonopy code \cite{r15} where finite displacement method was employed with a 2x2x1 supercell with a 4x4x1 K point mesh.
\par
The calculation of lattice thermal conductivity was executed in phono3py code \cite{r13} which is based on first principle Boltzmann transport equation (BTE)in the single mode relaxation time approximation with the same supercell and a 7x7x1 mesh.
Thermal conductivity is obtained by the relation
\begin{equation}
\kappa = \frac{1}{N V_0} \sum_{\lambda} C_{\lambda} \nu_{\lambda} \otimes \nu_{\lambda} \tau_{\lambda} ,
\end{equation}
where $V_{o}$ is the volume of the unit cell and $\nu_{\lambda}$ and $\tau_{\lambda}$ are the group velocity and phonon relaxation time respectively. $C_{\lambda}$ is the mode dependent heat capacity defined as
\begin{equation}
C_{\lambda}= k_{B} (\frac{\hbar \omega_\lambda}{k_{B} T})^{2} \frac{exp(\hbar \omega_\lambda)/k_{B}T}{[exp(\hbar \omega_\lambda/k_BT)-1]^{2}}.
\end{equation}
The eigen value equation directly gives the group velocity as per the equation 
$\nu_\alpha (\lambda) = \frac{\partial \omega_\lambda}{\partial q_{\alpha}}$.

\begin{figure}[ht]
	\begin{center}
		\includegraphics[scale=0.65]{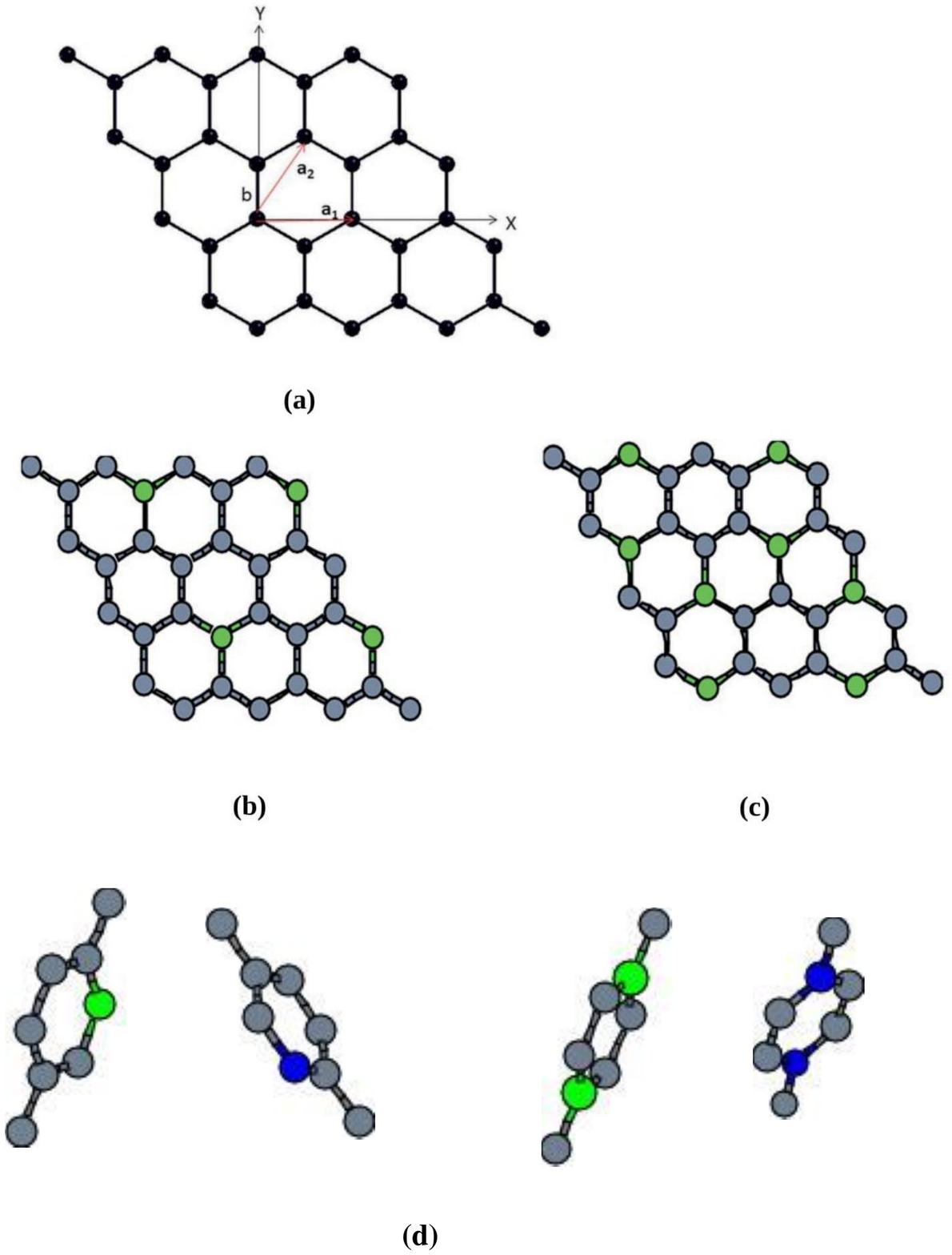}
	\end{center}
	\caption{Schematic representations (a) unit cell of Pristine Graphene where $a_1$ and $a_2$ represent the lattice vectors of unit cell and $b$ indicates the $C$-$C$ bond length i.e. 1.40 \AA, (b) 12.5 $\%$ $B$ or $N$ doped graphene Unit cell, (c) 25 $\%$ $B$ and $N$ doped graphene unit cells. The black/grey dots indicates $C$ atoms and green dots indicates the position of $B$ or $N$ atoms, (d) Side views of 8-atom unit cell with 12.5 $\%$ $B$ $\&$ $N$ and 25 $\%$ $B$ and $N$ doping }
\end{figure}

\section{Results and Discussions}
We started our investigation by considering the pristine graphene with a unit cell of 2 atoms per unit cell and 8 atoms per unit cell. The effect of B and N doping on thermal conductivity of graphene has been studied with 12.5 $\%$ and 25 $\%$ doping concentrations. The thermal conductivity of pristine and doped graphene have been calculated using methodology/computational details described above, and results are described in the following subsection:

\subsection{Pristine Graphene}
The graphene unit cell with minimum energy is illustrated in Figure-1(a) where $a_1$  and $a_2$  represent the lattice vectors of unit cell of 1.40 \AA.
A distance of approximately 10$\AA$  between two graphene sheets have been introduced artificially. The C-C distance is 1.42 \AA, which is consistent with the experimental value \cite{r16}. In a pristine graphene lattice with 2 atoms and 8 atoms in the unit cell, the dispersion relations have been calculated using Eq.(1) and shown in Figure-2(a) and 2(b). There are 6 phonon branches observed in the dispersion for graphene with 2 atom unit cell. In plane vibrations of graphene lattice are identified as linear transverse and longitudinal acoustic branches TA and LA, while out of the plane vibrations are termed as flexural phonon branches ZA and ZO. There are four allowed frequencies of a wave, namely longitudinal optical and transverse optical, longitudinal acoustical and transverse acoustical branches, respectively. The frequency of acoustic phonons reduces at longer wavelengths which are equivalent to sound waves in a lattice.
\par
The existence of flexural mode called the ZA mode is a basic feature in the phonon dispersion of graphene. ZA mode is the lowest frequency mode and most accessible to be excited.
Heat transport in graphene has been attributed to in-plane acoustic phonons as group velocities vanish for wave vector q = 0. The basis of emergence of this mode can be credited to the surface interactions. There is a quadratic variation in the ZA mode behavior near gamma point. Transverse motion produces ZA and ZO modes, but low- frequency ZA mode is of utmost importance and is more likely to be affected due to external influence. The flexural phonon has been found to contribute dominantly towards the lattice thermal conductivity of suspended graphene \cite{r61}. The phonon dispersion curves for pristine graphene matches well with the known experimental and other theoretical calculations \cite{r62}.
\par
We have calculated the thermal conductivity of pristine graphene using Eq.(2) in the temperature range of 100K-1000K. Thermal conductivity has been found to be 5000 W/mK, which decreases with the increase in the temperature, as shown in Figure-3(a). Thermal conductivity of graphene shows three distinct regions of variations with a sharp decrease exponentially upto 225 K, decrease as (1/T) upto 600 K and nearly constant at high temperature. The lattice thermal conductivity of graphene depends strongly on lattice vibration and mean free path of phonon, which increases exponentially with decreasing temperature. The large phonon means free path in the range of 100-600 nm have also been reported in supported and suspended graphene samples. The effect on lattice vibration has been investigated in terms of the phonon dispersion relation, as shown in Figure-2(a). The lattice specific heat and phonon density of states of pristine graphene is shown in Figure-4(a) $\&$ Figure-4(b). The calculated thermal conductivity values agrees well with the measured values and are consistent with temperature-dependent experimental results \cite{r6}.

\subsection{B doped Graphene}
For studying the effect of $B$ doping on the thermal conductivity of graphene, we have taken a unit cell with 8 $C$ atoms and doped $B$ with 12.5 $\%$ and 25 $\%$ doping concentrations, as shown in Figure-1 (b) and Figure-1(c).
The $B$ doping in graphene results in breaking of $P6/mmm$ hexagonal symmetry of pristine graphene. The lattice constant increases from 4.92 $\AA$ in pristine graphene to 5.04 $\AA$ and 5.17 $\AA$ for 12.5 $\%$ and 25$\%$ doping respectively. The results show a bond distance of 1.41 $\AA$ for $C$-$C$ bonds and 1.35 $\AA$ for $C$-$B$ bond distance in the lattice. The bond angles between $C$-$C$-$C$ and $C$-$B$-$C$ have been found to be 120.02$^o$ w.r.t 120$^o$ observed for pristine graphene.
\par
The $B$ dopant in graphene lattice interacts with $sp^2$ hydridization with $C$ atoms which results in change in the local electronic properties with stronger C-B bonds. As the B doping is increased, the electronic character of graphene changes from semi-metal to semiconductor. The pyramidalization angle for the doped graphene have been found to be zero. On doping graphene sheet with single B atom, the Fermi level notably shifts below Dirac point by 0.7 eV because of electron deficient character of boron atom. Bader charge analysis was used to calculate charge transfer that revealed a charge transfer of 1.79 e from boron to carbon.
Band gap of doped structure has been found to increase with the increase in $B$ doping concentration. The band structure of the $B$ doped graphene has been found to be deformed \cite{r1003}. 
\par
$B$-doping with $12.5\%$  concentration results in the elevation of the degenerate branches of the phonon frequencies resulting in separate branches in the phonon dispersion, as shown in Figure-2(c). There is a gradual softening of phonon branches with 25$\%$ doping. The frequency of the ZA mode has been found to decrease in both cases. The calculated thermal conductivity is illustrated in Figure-3(b) have shown a sharp decrease in value with B doping. The figure shows a decrease in the magnitude of thermal conductivity w.r.t pristine graphene to 190 W/mK for 12.5 $\%$ doping. The thermal conductivity decreases with an increase in the temperature, in a similar manner as observed for pristine graphene. The thermal conductivity decreases further to 30.0 W/mK for 25$\%$ doping concentration. The rate of change of thermal conductivity with temperature decreases significantly with increasing doping concentration.
The results are consistent with recent studies using Non-equilibrium molecular dynamics (NEMD), which has predicted that the addition of 0.75 $\%$ of $B$ atoms in graphene results in a decrease in thermal conductivity by more than 60 $\%$ \cite{r16}.

\begin{figure}[ht]
	\begin{center}
		\includegraphics[scale=0.65]{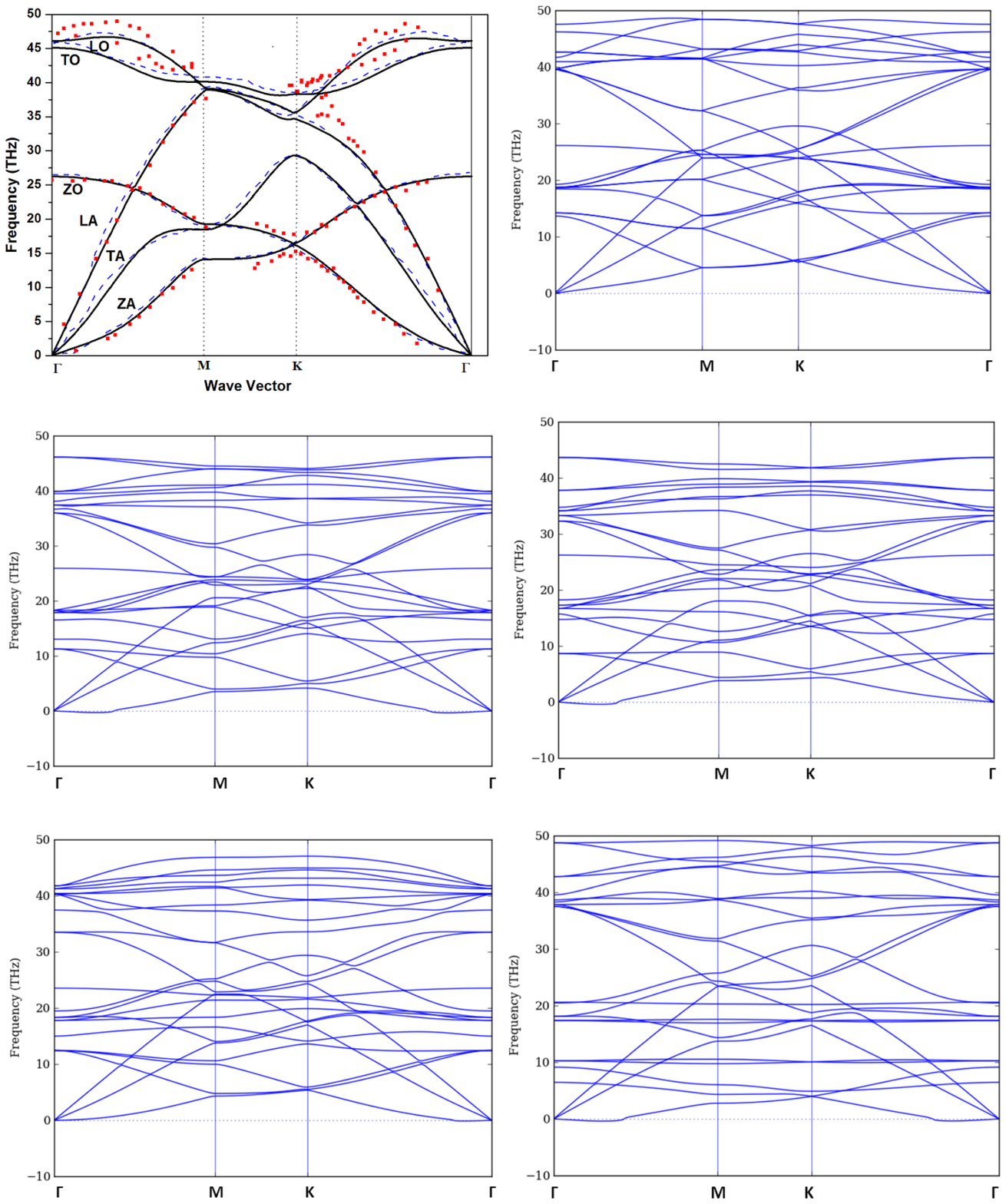}
	\end{center}
	\caption{Dispersion Relation of (a) Pristine Graphene with two atom unit cell. Dark lines indicate calculated values and dotted lines are experimental values (b) Pristine graphene with 8 atom Unit cell (c) $B$ doped graphene with 12.5 $\%$ doping, (d) $B$ doped graphene with 25 $\%$ doping, (e) $N$ doped graphene with 12.5 $\%$ doping and (f) $N$ doped graphene with 25 $\%$ doping.}
\end{figure}
\subsection{N doped Graphene}
In order to examine the effect of $N$ doping on thermal conductivity of graphene, we have considered a unit cell with 8 C atoms, doped $N$ with 12.5$\%$ and 25$\%$ doping concentrations as shown in Figure-1(c). The $N$ doping results in a change of the crystal symmetry of unit cell from $P6/mmm$ to $P1$ hexagonal symmetry in doped graphene. The lattice constant decreases from 4.92 $\AA$ in pristine graphene to 4.90 $\AA$ and 4.86 $\AA$ for 12.5 $\%$ and 25 $\%$ doping, respectively. $N$ also interacts through  $sp^2$ hybridization with the $C$ atom. The electron-rich character of the N , results in shifting of Fermi level above the Dirac point by 0.7 eV. The Bader charge analysis of $N$ doped graphene predicts a 1.16 e charge transfer from $C$ to $N$ .
The electronic properties of $N$ doped graphene is similar to B doped graphene, except the shift in the $E_F$ energy above the Dirac point \citep{r1003}.
\par
The change in phonon dispersion relation of graphene on $N$ doping concentration with 12.5 $\%$ and 25 $\%$ doping has been plotted in the Figure-2(e) and Figure-2(f). Figure shows that similar to $B$ doping, $N$ doping also results in lifting the degeneracy of pristine graphene even with 12.5 $\%$ doping. For 25 $\%$ doped structure, the softening of acoustic and optical phonon branches is observed in phonon dispersion curve.
\par

The calculated thermal conductivity of the N doped graphene as a function of temperature is shown in Figure-3(c). The figure shows a sharp decrease in the thermal conductivity to 4 W/mK with 12.5 $\%$ doping at low temperature and remains constant near zero for higher temperatures. The thermal conductivity for 25 $\%$ doping concentration has shown interesting behavior w.r.t to 12.5 $\%$ doping level. The thermal conductivity increases to order of 55 W/mK, which is contrary to the behavior of B doped graphene, which have exhibited inverse relation with higher doping concentration. The rate of change of thermal conductivity with temperature changes significantly in different temperature ranges: thermal conductivity decreases sharply for low temperature below 200 K, value increase marginally between 200K and 400K, and decrease linearly over 400 K.
\par
The results are consistent with the recent studies using equilibrium molecular dynamics based on Green Kubo method, which has predicted a drastic decline in the thermal conductivity of N doped graphene even for a very low concentration of 0.5$\%$ and 1.0 $\%$. The results are consistent with the recent studies using non-equilibrium molecular dynamics (NEMD), which has predicted that the addition of even 1 $\%$ N doping in graphene results in a decrease of thermal conductivity by more than 50 $\%$ \cite{r17}. 

\begin{figure}[ht]
	\begin{center}
		\includegraphics[scale=0.65]{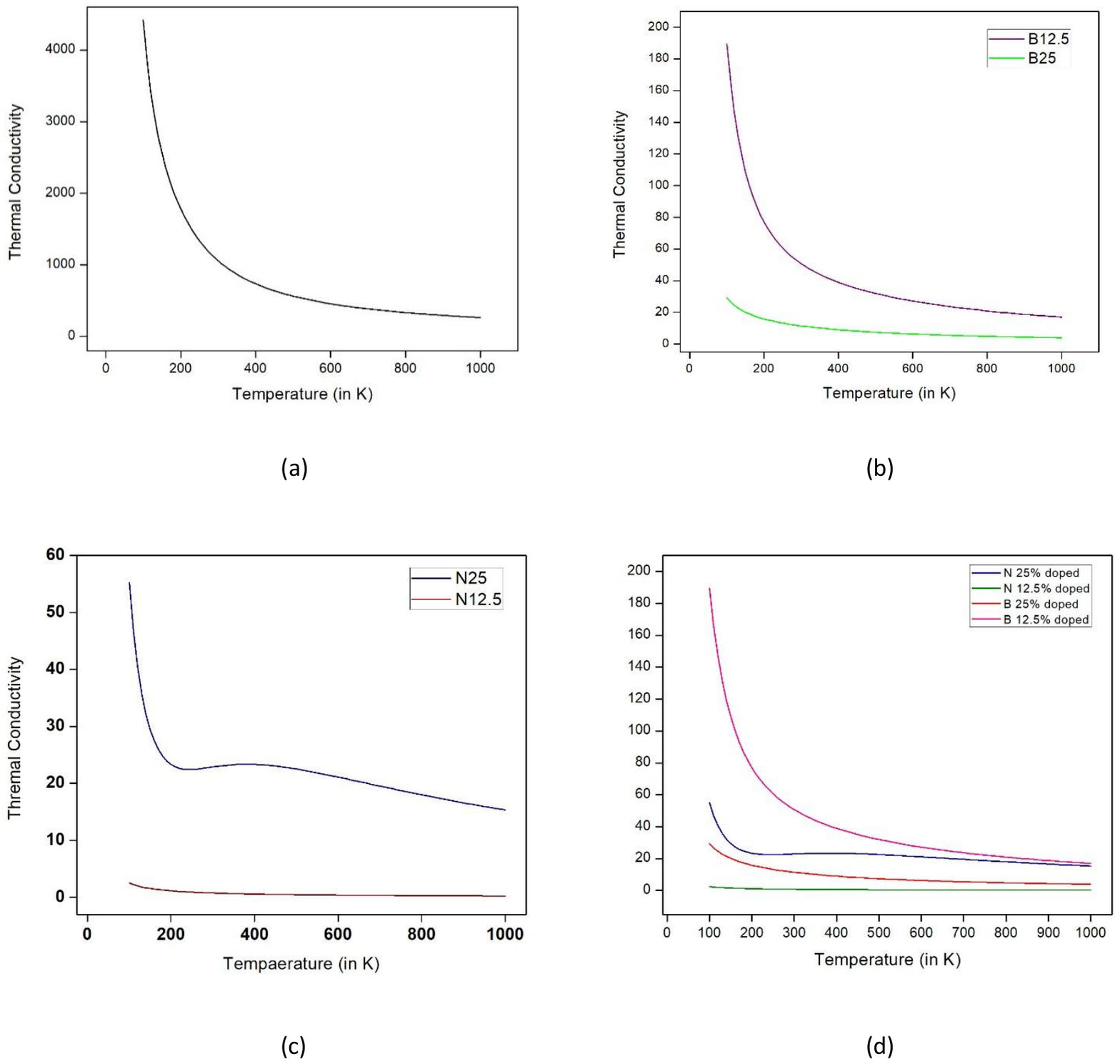}
	\end{center}
	\caption{Thermal Conductivity (W/mK) as a function of temperature of  (a) pristine graphene, (b) $B$-doped graphene with 12.5 and 25 $\%$ doping, (c) $N$-doped graphene with 12.5 and 25 $\%$ doping and (d) Comparative values of thermal conductivity of pristine and all doped samples.}
\end{figure}

\par
The reduction in the lattice thermal conductivity in the lattice thermal conductivity of pristine graphene on doping with B and N can be attributed to the change in the phonon related properties such as a change in the phonon dispersion relation, group velocity,lattice specific heat capacity and relaxation time. The presence of dopants presents an obstacle for the phonon propagation through the graphene lattice and changes the dispersion relation, as shown in Figure-2. As the doping concentration is increased, the dispersion branches start ungrouping, which leads to non-degenerate modes and further increasing the possible scattering phenomena. We can say that for doped graphene, the non-degenerate phonon modes largely govern the scattering phenomena, limiting the phonon mean free paths that reduces the thermal conductivity. In graphene, large part of contribution to thermal conductivity is coming from low-frequency flexural phonons and has shown significant change with change in doping concentration. 
\par
The doping of B-N in graphene probably induces phonon scattering, which is expected to  change the intrinsic phonon mean free path and phonon density of states (PDOS). Figure 4 shows the calculated values of lattice specific heat and PDOS of doped graphene. The lattice specific heat for pristine graphene is plotted in Figure 4 (a) and it is in agreement with the known theoretical and experimental results \cite{r17a,r18,r19,r20}.
The specific heat has a higher magnitude at low temperature, which can be described on the basis of higher PDOS at low frequency. Higher PDOS leads to the large activation of low-frequency phonons at lower temperatures, and hence, at room temperature, large phonon-phonon scattering occurs for doped graphene.

As a result, thermal conductivity of doped graphene is lower than pristine graphene at any particular temperature. When doping concentration is increased, then there is a shift in density of states away from the graphene limit, and an increase is observed in the phonon-phonon scattering intensity of low-frequency phonons,that leads to an overall decrease in thermal conductivity. Therefore, the change in the scattering interaction modes due to doping, alongwith interaction modes in pristine graphene, is a primary source of reduction in the thermal conductivity of doped graphene. The change in the phonon dispersion relation of doped graphene results in increase in scattering process that gives intermediate frequency phonons for Umklapp scattering phenomenon.

\begin{figure}[ht]
\centering
\includegraphics[scale=0.62]{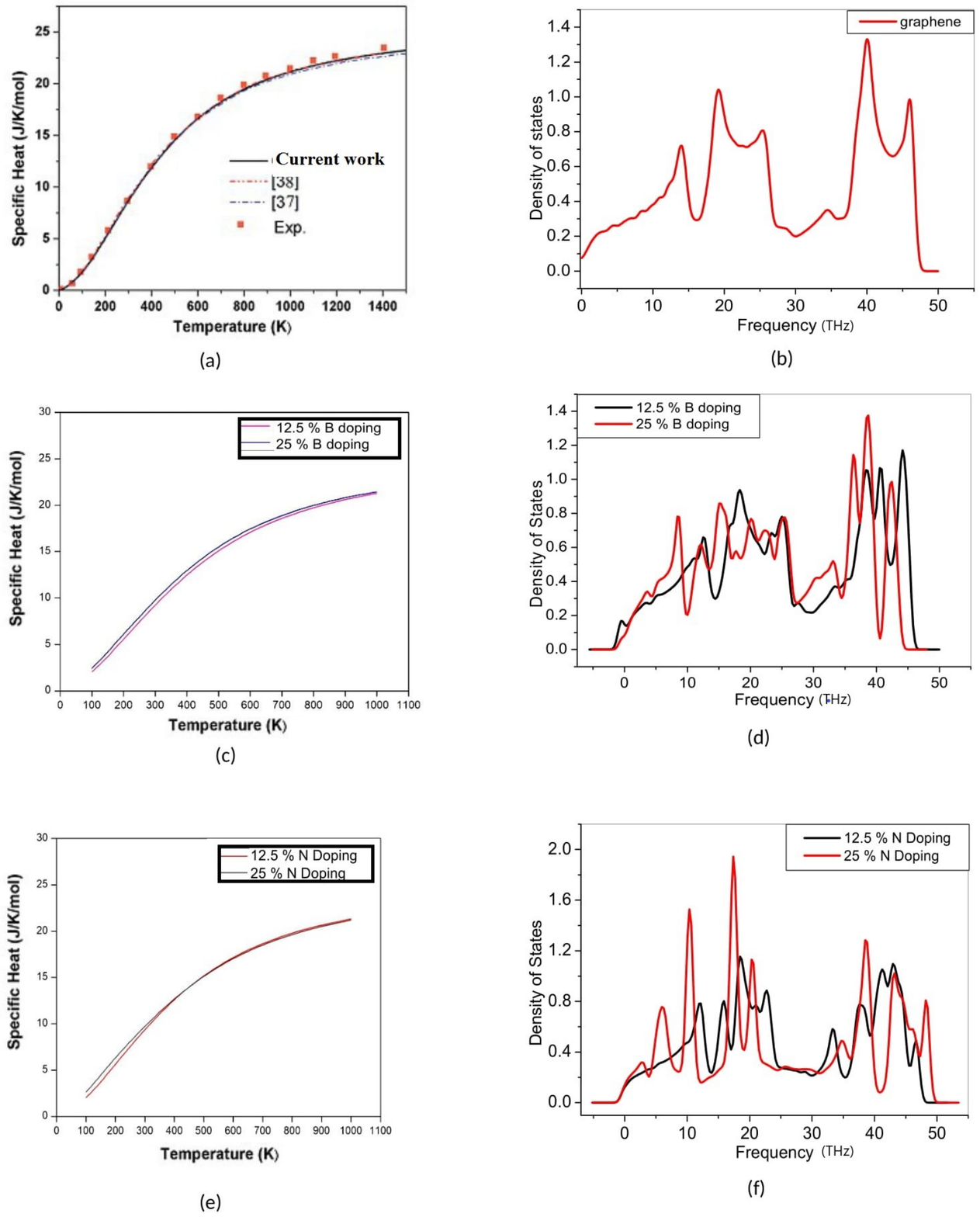}
\caption{(a) The lattice specific heat pristine graphene, dotted line are experimental results \cite{r17a} (b) phonon density of states of pristine graphene (c) lattice specific heat of $B$ doped graphene with 12.5 $\%$ and 25 $\%$  (d) Phonon Density of states of $B$ doped graphene with 12.5 $\%$ and 25 $\%$ doping (e) lattice specific heat of $N$ doped graphene with 12.5 $\%$ and 25 $\%$  (f) Phonon Density of States of $N$ doped graphene with 12.5 $\%$ and 25 $\%$ doping.}
\label{fig:specific heat}
\end{figure}

\section{Summary and Conclusions}
We have used density functional theory calculations using plane wave basis VASP code in combination with phono3py for calculation of thermal conductivity of pristine and B / N doped graphene. The anharmonic effects were included by considering cubic force constants in the calculations. The results have shown that pristine graphene possess high thermal conductivity of order 5000 W/mK at low temperature, and its magnitude decreases with an increase in the temperature. The thermal conductivity decreases sharply when graphene is doped with B and N. The thermal conductivity of B doped graphene has been found to be 190 W/mK and 30 W/mK for 12.5 $\%$ and 25 $\%$ doping concentration at low temperature. N doped graphene shows thermal conductivity of  the order of 4 W/mK and 55 W/mK for 12.5 $\%$ and 25 $\%$ doping concentration, respectively. The B doped graphene had shown higher thermal conductivity in comparison to the N doped graphene for 12.5 $\%$ doping concentration, whereas the trend is reversed when doping concentration increased to 25$\%$. The thermal conductivity of N doped graphene has been found to be less sensitive to change in the temperature in comparison to the pristine graphene and B doped graphene.

The sharp decrease in the thermal conductivity in doped graphene can be attributed to impurity induced phonon scattering, change in the lattice vibration frequency, and phonon relaxation time. However, the results of doped graphene could not be verified due to the lack of experimental results on doped graphene. These results have highlighted the need for carrying out more experimental investigations into doped graphene with different levels of B and N doping concentration. We believe without results of thermal conductivity of doped graphene, the earlier study of electronic properties alone was incomplete and of not much practical utility. The results have highlighted the feasibility of tuning the thermal conductivity of graphene by doping for thermoelectric applications where minimum lattice thermal conductivity is preferred.

\section*{Conflicts of interest}
There are no conflicts of interest to declare.
		
\pagebreak

\end{document}